\documentclass[useAMS,usenatbib,epsfig, a4paper]{mn2e}

\usepackage{epsfig}
\usepackage{times}
\usepackage{amssymb}

\newcommand{\ETAL}{{ et al.}}
\newcommand{\be}{\begin{equation}}
\newcommand{\ee}{\end{equation}}
\newcommand{\ba}{\begin{eqnarray}}
\newcommand{\ea}{\end{eqnarray}}

\def\gtorder{\mathrel{\raise.3ex\hbox{$>$}\mkern-14mu
             \lower0.6ex\hbox{$\sim$}}}
\def\ltorder{\mathrel{\raise.3ex\hbox{$<$}\mkern-14mu
	                  \lower0.6ex\hbox{$\sim$}}}

\title[Fast and reliable MCMC]
    {Fast and reliable MCMC for cosmological parameter estimation}
\author[J.~Dunkley et al.]
{Joanna Dunkley$^{1}$, Martin Bucher$^{2,4}$, Pedro G. Ferreira$^{1,4}$, 
Kavilan Moodley$^{1,3,4}$, and
\newauthor 
Constantinos Skordis$^{1}$\\
$^{1}$Astrophysics, University of Oxford, Denys Wilkinson Building, 1 Keble 
Road, Oxford OX1 3RH, United Kingdom \\
$^{2}$DAMTP, Centre for Mathematical Sciences, University of Cambridge,
Wilberforce Road, Cambridge CB3 0WA, United Kingdom \\
$^{3}$School of Mathematical Sciences, University of KwaZulu-Natal, Durban, 
4041, South Africa \\
$^{4}$African Institute for Mathematical Sciences (AIMS), 6-8 Melrose Road, 
Muizenberg 7945, South Africa \\
}
\date{\today}
\pubyear{2004}

\begin{document}

\maketitle

\begin{abstract}
Markov Chain Monte Carlo (MCMC) techniques are now widely used for cosmological
parameter estimation. Chains are generated to sample the posterior 
probability
distribution obtained following the Bayesian approach.
An important issue is how to optimize the efficiency of such sampling and 
how to diagnose whether a finite-length chain has adequately sampled
the underlying posterior probability distribution.
We show how the power spectrum of a single such
finite chain may be used as a convergence diagnostic by means of a fitting
function, and discuss strategies for optimizing the distribution for
the proposed steps. 
The methods developed are applied to current CMB and LSS data interpreted using
both a pure adiabatic cosmological model and a mixed adiabatic/isocurvature
cosmological model including possible correlations between modes. For 
the latter application,
because of the increased dimensionality and the presence of degeneracies,
the need for tuning MCMC methods for maximum efficiency becomes 
particularly acute.

\end{abstract}


\section{Introduction}
\label{sec:intro}

The availability of high quality data from both CMB (Hinshaw et al.; 
Kogut et al. 2003) and 
large-scale structure (Percival et al. 2001; Tegmark et al. 2003) experiments 
has allowed the field of precision cosmology to advance rapidly
in the last few years. 
Methods of cosmological parameter estimation are allowing us to narrow down 
the range of possible universes by placing bounds on the 
parameters describing a particular model. 
Because many of the models have a large number of parameters, 
ranging in complexity from the simplest 
scale-invariant $\Lambda$CDM cosmology to those including spatial curvature, 
massive neutrinos, a dark energy equation of state, cosmic strings, or a 
non-adiabatic contribution,
it has become commonplace to use Markov Chain Monte Carlo (MCMC) methods
to explore the probability distributions of high 
dimensionality obtained following
Bayesian statistics.

MCMC methods were first introduced in the 1950s 
(Metropolis et al. 1953) to sample
an unknown probability distribution efficiently
and are described in detail 
in Neal (1993) and in Gilks, Richardson \& Spiegelhalter (1995).
Instead of calculating the probability density at sites on a 
regular grid spanning the entire parameter space, one draws samples
sequentially according to a probabilistic algorithm. 
The sequence of states visited forms a Markov Chain distributed
according to the probability distribution to be explored. 
Rather than scaling exponentially with the number of 
parameters varied, the time needed to sample a distribution grows 
approximately linearly with 
dimension. For this reason MCMC methods are particularly useful 
for cosmological parameter estimation. This application has 
explicitly been discussed by
Christensen et al. (2001), Knox, Christensen \&  Skordis (2001), 
Lewis \& Bridle (2002),  Verde et al.~(2003), Tegmark et al.~(2003)
and others.

All MCMC algorithms share the property that asymptotically 
the distribution of 
states visited by the chain is identical to the underlying 
distribution. However, it is critical to 
be able to determine whether and to what extent
a finite length chain is a fair sample and can be used 
to make confident and accurate estimates of statistics characterizing
the underlying distribution. This idea 
of `convergence' to a stationary 
distribution has been discussed extensively in the statistical literature 
(see e.g. Cowles \& Carlin 1996 for a review), and there exist a range 
of convergence tests that can be applied to 
MCMC output, many of which are described in the {\sc coda} manual (Best, 
Cowles \& Vines 1995).
Both Heidelberger \& Welch (1981,1985) and Geweke (1992) use spectral 
methods to determine 
the total running length of a simulation and the length of an initial 
transient to be discarded.  Gelman \& Rubin (1992) study the dispersion 
of the means of multiple chains, while Raftery \& Lewis (1992) use second 
order Markov chains to insure that percentiles are estimated within a 
given accuracy. 
In the cosmological context, the package of {\sc coda} 
tests were first applied 
by Christensen et al. (2001), with the Gelman \& Rubin test further explored 
by Verde et al. (2003) and included with the Raftery \& Lewis test in 
the {\sc cosmomc} package (Lewis \& Bridle 2002).  Methods for 
speeding up the convergence have 
been considered by Slosar \& Hobson (2003), in the 
{\sc cosmomc} package and by Doran \& Mueller (2003).

Here we describe a convergence test for MCMC methods 
that we have tested and found to work well for several 
CMB parameter estimation problems.
The test tells us when one can stop 
running the MCMC chain and use it to estimate
statistics of the underlying distribution. 
We draw upon various techniques incorporated in existing convergence 
analyses, studying
the spectral behaviour generic to chains generated with the 
Metropolis algorithm. 

This article is organized as follows. Section 2 reviews the 
Metropolis algorithm and section 3 defines the convergence
of a chain. Section 4 formulates a spectral convergence 
test based on fitting to a template, which we have empirically
found to work well, to the sample power spectra of finite
length chains. This section demonstrates the effectiveness
of the proposed spectral test using Gaussian distributions
of various dimensions and differing ratios of the covariance
of the trial distribution to that of the underlying
distribution. Section 5 explores how to optimize the 
trial distribution for the proposed steps, tailoring
its covariance to that of the underlying distribution.
In section 6 we investigate what can go wrong, considering
various non-Gaussian distributions, ranging from 
mildly non-Gaussian to more pathological. Section 7
summarizes our method in the form of an explicit 
`recipe,' which is then applied to CMB+LSS data
in section 8. Finally, we conclude with a discussion in
section 9.  

\section{The Metropolis Algorithm} 
\label{sec:met}

The Metropolis algorithm, used to sample a probability 
distribution $p(x)$ in one dimension, works as follows. Starting at
an initial position $x_0$, we generate a sequence of points 
$x_1, x_2,\ldots $ according to the following rule. 
$x_{i+1}$ is generated from $x_i$ by 
attempting a trial step $x_{trial}$  
distributed according to a trial distribution 
$q_{trial}(x_{trial},x_{i}),$ typically but not always chosen
with the special form 
$q_{trial}(x_{trial},x_{i})=q_{trial}(x_{trial}-x_{i}).$
The trial distribution must be symmetric
and chosen such that all points with
$p(x) \ne 0$ can be connected by the chain.
Because of considerations of detailed balance, the 
symmetry of $q_{trial}$ ensures that $p(x)$ is stationary
under the Markov process. 
We shall typically use for $q_{trial}(x_{trial}-x_{i})$ 
a Gaussian of vanishing mean and adjustable 
variance $\sigma^2 _T.$ 
If $P_R = p(x_{trial})/p(x_i) \ge 1$, then $x_{i+1}$ is 
set to $x_{trial}$ with probability one. Otherwise, $x_{i+1}$ is set to 
$x_{trial}$ with probability $P_R$ and to $x_i$ with probability 
$(1-P_R)$. If the chain moves to 
$x_{trial},$ the step has been `accepted'. Otherwise, we say that it has 
been rejected. 
The correlated chain of steps $x_i$ explores the full 
range of the sample space spanned by $p(x)$ such that eventually, in
the infinitely long chain limit, the distribution of points
visited is exactly described by the 
distribution $p(x)$. Most of the time is spent exploring
regions of high likelihood, but all regions where $p(x)$ is non-zero
are eventually explored.

To sample a $D$-dimensional distribution $p({\bmath x})$ using the 
Metropolis algorithm, $x_{trial}$ 
is replaced by ${\bmath x}_{trial}$, where 
$(\bmath{x}_{trial}-\bmath{x}_i)$ is distributed according to 
a multivariate Gaussian of zero mean and 
covariance matrix ${\bmath{C}}_T$. This $D$-dimensional step 
${\bmath{x}}_{trial}$ can be drawn from a 
Gaussian distribution of non-diagonal covariance
by drawing $D$ Gaussian random samples ${\bmath{y}}$ with 
unit variance and generating ${\bmath{x}}_{trial}=
{\bmath C}^{-1/2}_T{\bmath{y}},$ where ${\bmath C}^{-1/2}_T$ is the positive 
definite matrix square root.

\section{Defining Convergence}
\label{sec:define}

The distribution of points visited by the infinite MCMC chain described
above is described exactly by the underlying probability density 
$p(x)$. Statistical properties of the distribution such 
as the mean, median, and quantiles can 
therefore be calculated directly from the chain.
In practice, one must use a chain of finite length to 
estimate these statistical properties of the underlying distribution.
Errors arise from the truncation to a finite chain both because of
shot noise and because of correlations between successive elements of
the chain. Following the literature, we shall say that a finite chain 
has `converged' 
when its statistical properties, suitably defined, reflect those of the 
underlying distribution $p(x)$ with `sufficient accuracy' that the
chain can be terminated. 
To determine whether convergence has been achieved requires 
answering the following two questions:

\vskip 4pt
\noindent
1. Has the chain fully traversed the region of high probability in such a way 
that the correlations between successive elements of the chain
does not bias the inferred distribution for $p(x)$? Apart
from shot noise, a chain that is too short may explore only a single
peak or portion of a single peak where $p(x)$ is significant.

\vskip 4pt
\noindent
2. Can we then estimate suitably defined statistics 
about the underlying distribution $p(x)$ with sufficient accuracy?

\vskip 4pt

In the next Section we show how to extract from a single chain 
the information about the large-scale correlations of the chain that 
indicate whether the first requirement above has been satisfied. 
To address the second question, a level of accuracy for a given
statistic must be specified.  Let $\mu$ and $\sigma^2_{0}$ be the mean and 
variance of the underlying distribution $p(x),$ respectively.
A useful diagnostic is the variation of the sample means $\bar x$
obtained from a finite chain. Let $\sigma_{\bar{x}}^2(N)$ 
indicate the variance of the sample mean, defined by averaging over
independent realizations of a finite chain of length $N.$ We assume 
a starting point determined according to the underlying distribution,
in other words that any initial transients of the chain have decayed away.   
The variance in the sample mean may be characterized by the dimensionless
ratio
\be
r={\sigma_{\bar{x}}^2\over \sigma_{0}^2}, 
\ee
which we shall call the `convergence ratio'. We require $r$ to lie below 
some cutoff value, chosen for example to be $0.01$.
The sample mean variance has been used in various ways in the literature as
a diagnostic of convergence, as outlined in Cowles \& Carlin (1994).
Heidelberger \& Welch calculate confidence intervals on the mean using the 
ratio $\sigma_{\bar{x}}/{\bar{x}}$, but do not use the distribution 
variance. The Gelman \& Rubin test incorporates a 
similar ratio: their $R$ statistic roughly translates to $R \sim 1+r$, but 
the quantity is calculated using multiple parallel chains.
In the next section we show how to estimate the
convergence ratio $r$ from a single chain.

If a chain is started far outside the region of high probability, an 
initial section of the chain consisting of steady progression into
this region will be unrepresentative of 
the underlying distribution and must be discarded.  In the literature
this initial transient has been dubbed `burn-in.' 
We use the ratio of the probability $p$
compared to the maximum $p_{max}$ as an indicator of how 
much of the beginning of the chain to chop out, for example while 
$p/p_{max} < 0.1$. Such truncation 
is unnecessary if the chain is started from a point already 
known to lie in the region of high probability, for example from an
earlier chain.

\section{Spectral analysis of MCMC chains}
\label{sec:spectrum}

An infinite one-dimensional Gaussian random chain for which the 
underlying statistical
process is independent of position may be expanded into Fourier 
coefficients according to 
\be
x_n=\int _{-\pi }^{+\pi }{dk\over \sqrt{2\pi }}~\tilde x(k)~e^{ikn},
\ee
where the reality of the $x_n$ implies that
$\tilde x(k)=\tilde x^*(-k).$ The two-point correlations can
be characterized in terms of the Fourier coefficients 
\be
\langle \tilde x(k)~\tilde x^*(k')\rangle =\delta (k-k')~P(k),
\ee
where the power spectrum $P(k)$ is an even function. Any correlation 
linking unequal $k$ and $k'$ would contradict the 
hypothesis of position independence. 

We may define the two-point autocorrelation 
(i.e., the chain variance) 
\be
C_0=\langle x_n^2\rangle =\int  _{-\pi }^{+\pi }{dk\over 2\pi }~P(k),
\ee
giving an average measure of the power spectrum. Similarly, the correlation 
with an offset $N$ is given by
\be
C_N=\langle x_n~x_{n+N}\rangle =\int  _{-\pi }^{+\pi }{dk\over 2\pi }~P(k)~
\cos (kN).
\ee
We consider the sample mean
\ba
{\bar x}_N={1\over N}\sum _{n=0}^{N-1}x_n,
\ea
the variance of which is given by
\ba 
\left< {{\bar x}_N}^2\right> ={1\over N}\int _{-\pi }^{+\pi }{dk\over 2\pi N}~
{\sin ^2[Nk/2]\over \sin ^2[k/2]}~P(k)
\label{zzz}
\ea
for a chain with zero mean. Since for all integers $N>0,$ 
\ba 
\int _{-\pi }^{+\pi }{dk\over 2\pi N}~
{\sin ^2[Nk/2]\over \sin ^2[k/2]}=1,
\ea
eqn.~(\ref{zzz}) is a weighted average of $P(k)$ that becomes more and more
concentrated around $k=0$ as $N$ becomes large. 
Because
\ba
\lim _{N\to \infty }{1\over 2\pi N}{\sin ^2[Nk/2]\over \sin ^2[k/2]}=\delta (k),
\ea
it follows that for large $N$
\ba
\sigma^2_{\bar{x}}=\left< {{\bar x}_N}^2\right> \approx {1\over N}\cdot P(k=0).
\ea
Consequently, estimating the sample mean variance of a long chain is 
equivalent to
estimating $P(k)$ at $k=0.$

In practice, we want to estimate the power spectrum $P(k)$ from a finite 
chain of length $N$. To do so we define the derived random variables
\ba
a_N^j={1\over \sqrt{N}}\sum _{n=0}^{N-1} x_n~\exp \bigl[ i2\pi (jn/N)\Bigr] 
\ea
where 
$[j=-(N/2-1),$ $-(N/2-2),$ $\ldots ,$ $-1,$ $0,$ $+1,$ $\ldots ,$ $+(N/2-1),$ 
$+(N/2)]$ 
and $N$ is even and considered fixed. These coefficients are the discrete 
Fourier transform of the chain.  The $N$ new variables $a_N^j$ 
result from a unitary transformation acting on the original chain 
and are Gaussian and independently distributed, with variance
\ba
\left< \vert a_N^j \vert ^2 \right> =
{1\over N}\int _{-\pi }^{+\pi }{dk\over 2\pi N}~
{\sin ^2[N(k-\bar k_j)/2]\over \sin ^2[(k-\bar k_j)/2]}~P(k)\approx
P(\bar k_j)
\label{zzb}
\ea
where $\bar k_j=2\pi (j/N).$ 

The variables $a_N^j$ can be used to estimate the 
power spectrum
$P(\bar k_j)$ from a single realisation of a finite length chain.
We first calculate the discrete power spectrum ${\hat P}_j$ of the 
single finite chain
\ba
{\hat P}_j=\vert a_N^{+j}\vert ^2.
\label{cj:def}
\ea
for $j=1,\ldots. N/2-1.$ 

Adopting the approximation on the far right of 
Eqn.~(\ref{zzb}) as exact,
we find that ${\hat P}_j/P(\bar k_j)$ obeys a $\chi ^2$ distribution
with two degrees of freedom. 
We find that
\ba 
\left< \ln[{\hat P}_j]\right> = \ln [P(\bar k_j)] - \ln[2] + {\cal A}_1
\label{hatp}
\ea 
and 
\ba
{\rm Var}(\ln[\hat P_j])= 
{\cal A}_2-({\cal A}_1)^2={\pi ^2\over 6}\approx 1.645 
\ea
where
\ba 
{\cal A}_1={1\over2}\int _0^\infty dx~ \ln [x]~\exp [-x/2]
          =\ln [2]-\gamma \approx 0.1159
\ea
and 
\ba
{\cal A}_2={1\over 2}\int _0^\infty dx~ \ln ^2[x]~\exp [-x/2]
          ={\pi ^2\over 6}+(\ln [2]-\gamma )^2
\ea
where $\gamma \approx 0.577216$ is the Euler-Mascheroni constant.

$P({\bar k}_j)$ can therefore be inferred by fitting $\ln [{\hat P}_j]$ to 
template power spectra using least squares, providing an 
estimate for $P(k=0).$

\subsection{MCMC Power Spectra}

\begin{figure}
\epsfig{file=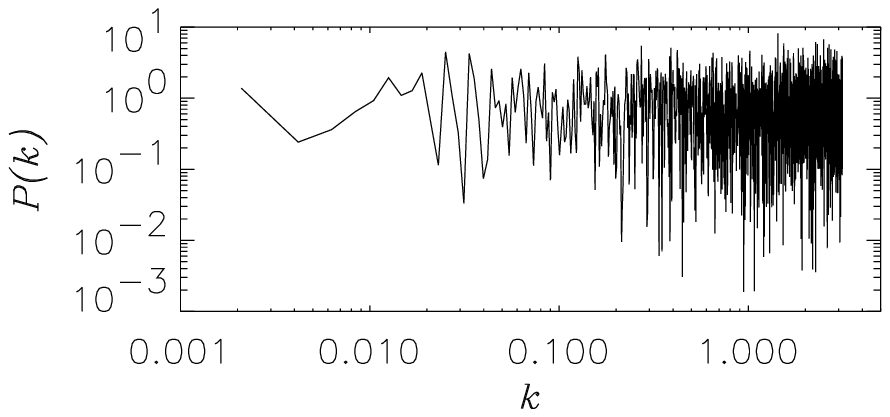,width=8.5cm,height=3.5cm}
\epsfig{file=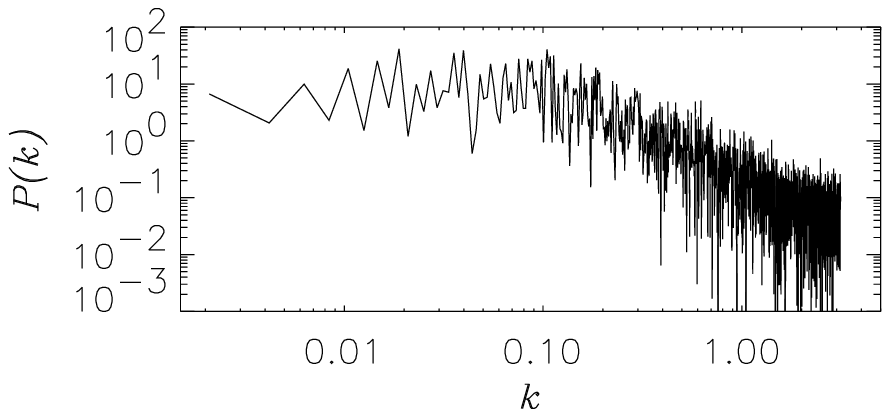,width=8.5cm,height=3.5cm}
\caption{(Top) The discrete power spectrum of an `ideal,' uncorrelated 
chain formed 
by drawing points at random from a Gaussian distribution of
unit variance. (Bottom) The discrete
power spectrum from an MCMC chain of length 
$N=3000$ sampling the same 
five-dimensional Gaussian.}
\label{ideal}
\end{figure}

An ideal sampler draws from the underlying distribution
with no correlations between
successive elements of the chain. The power spectrum is absolutely
flat, with $P(k)=\sigma ^2$ for all $k.$ Here $\sigma $ is the standard
deviation of the underlying distribution. This is a white noise spectrum.
Fig.~\ref{ideal} plots the ${\hat P}_j$ obtained from such a chain.
In this case, satisfying the convergence criterion $r < 0.01,$ 
requires only 100 steps.

An actual MCMC chain always has correlations on small scales due to 
the nature of the Metropolis algorithm. If the trial distribution 
is chosen such that very small trial steps are attempted, the chain 
propagates diffusively, behaving locally like a random walk. For large 
trial steps, on the other hand, the chain remains stuck at one point for 
quite some time before accepting a step. The initial and final points
of these jumps are almost uncorrelated, but the jumps occur infrequently. 
A balance between these two extremes through
a judicious choice of the trial distribution minimizes 
the correlations between successive elements of the chain.
Once the chain has fully travelled around the region of high probability,
the correlations at the largest scales
begin to vanish, and the large-scale behaviour starts to mimic an 
ideal sampler.
 
\begin{figure}
\epsfig{file=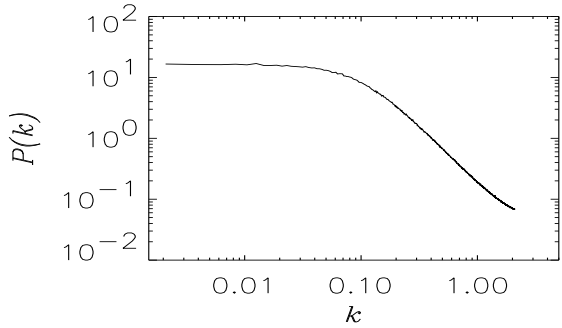,width=8.5cm,height=5cm}
\caption{The exact power spectrum of one of the variables of 
an MCMC chain sampling the same 5-D Gaussian distribution 
as in the lower panel of Fig. \ref{ideal},
measured by averaging over a large number of chains.}
\label{model}
\end{figure}

If we examine the power spectrum of an actual chain, we observe 
a white noise spectrum on large scales where $k$ is small,
turning over to a spectrum with suppressed power at large $k.$ The position of
the knee where the white noise turns over to suppressed power
with a different power law reflects the inverse correlation 
length. We illustrate this behaviour
with an MCMC chain 
sampling a five-dimensional 
Gaussian model with 
zero mean and unit variance in each dimension, sampled with an MCMC chain of 
length $N=3000$ using the Metropolis algorithm. The trial
distribution is a Gaussian of width $\sigma_T=1.1$, 
which we later show to
be optimal. The five-dimensional chain better illustrates
the effect of correlations on the power spectrum 
than a less correlated one-dimensional chain.
The power spectrum of one of the components of the chain 
is shown in the lower panel of Fig. \ref{ideal}. 
To remove the noise we simulate 5000 identical chains 
and find the average of the sample spectra as
shown in Fig. \ref{model}. Except at the 
very smallest scales, the small-scale behaviour is well approximated
by a power-law of the form $P(k) \propto k^{-\alpha}$,
with $\alpha $ typically in the neighbourhood of but not exactly equal
to 2, which would correspond to the spectrum of a perfect random walk. 

\subsection{A parametric fit to the power spectrum}

In analysing the power spectrum of a chain of finite length, we would
like to answer two questions: (1) Whether the white noise part of the 
spectrum has been reached for the lowest accessible $k$ (i.e.,
$k=j(2\pi /N)$  where $j$ is a small nonzero positive integer). 
Note that an estimate at $k=0$ is not accessible because
we almost always lack independent knowledge of the mean of the underlying
distribution. (2). What is our estimate for $P(k=0)$ so that
the variance of the sample mean can be estimated? 

Question (1) is not really a problem involving hypothesis testing---that is,
testing the hypothesis that the spectrum is in fact exactly white 
noise---because at the outset we know 
that at any finite $k$ the spectrum is not exactly flat. 
Instead the relevant
question is whether one is probing $k$ small enough so that the asymptotic
deviation from white noise present on the largest scales is sufficiently
small. Because of this, what we really want to do is fit to a template
that models the transition from white noise
at the largest accessible scales to correlated 
behaviour on small scales. 

An appropriate template has the following form:
\be{
P(k)=P_0\frac{(k^*/k)^\alpha}{(k^*/k)^\alpha+1}}.
\label{template:e}
\ee
Here the free parameter $P_0$ gives the amplitude of the white
noise spectrum in the $k\to 0$ limit. The parameter $k^*$
indicates the position of the turnover to a different power law
behaviour, characterized by the free parameter $\alpha ,$ at
large $k.$
This template models the observed behaviour 
more closely than for example a polynomial fit used by 
Heidelberger \& Welch (1981) to estimate $P(k=0)$ from the power spectra of 
discrete event simulations.

For a chain of length $N,$
our power spectrum data consists of a single
realization of the random variables
${\hat P}_j,$ $j=1,\ldots ,N/2-1,$ 
defined in eqn. (\ref{cj:def}).
For the parametric model defined by the above template, 
\ba
\ln {\hat P}_j&=&\ln [P(\bar k_j)]-\ln[2]+{\cal A}_1+r_j\cr
              &=&\ln [P_0]+\ln \left[
                   {(Nk^*/2\pi j)^\alpha \over 1+(Nk^*/2\pi j)^\alpha }
                    \right] -\gamma + r_j
\ea
where 
$r_j$ are random measurement errors 
characterized by
the expectation values 
$\langle r_i\rangle =0,$
$\langle r_ir_j\rangle =\delta _{ij}\pi ^2/6.$
We fit $\ln P_0, k^*$ and $\alpha$ 
using least squares, over the range of Fourier modes 
$1\le j\le j_{max}$.  
For a spectrum which turns over at $j^*=k^*(N/2\pi)$,
an appropriate limit is $j_{max}\sim 10j^*$, which gives 
equal weighting to the two power law regimes in log space 
and avoids using the very highest $j$ values which have 
small scale artefacts.
In practice the limit $j_{max}=1000$ is typically used
as a starting point, but a second iteration may be 
used to get a better fit once $j^*$ is known. The results are
fairly insensitive to the exact value, but we find 
$P_0$ to be overestimated when $j_{max} \gg 10j^*$.

\begin{figure}
\epsfig{file=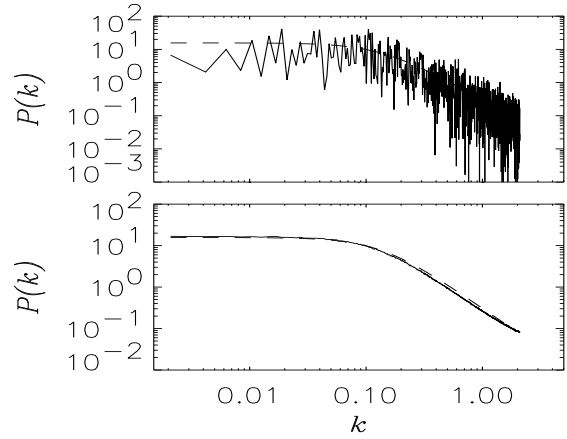,width=8.5cm,height=5cm}
\vskip 0.25in
\caption{
In the top panel, the chain of the lower panel in Fig.~(\ref{ideal})
is fitted to the template defined in Eqn.~(\ref{template:e}).
This fit is indicated by the dashed curve. The lower panel 
compares the obtained fit to the measured exact power spectrum
(shown as the solid curve).
}
\label{fit}
\end{figure}

We now test how well the functional form of this template 
fits the exact power spectra of the MCMC chains. We fit the 
template initially to the ${\hat P}_j$
of a single chain of length $N=3000$, as shown in Fig. \ref{fit}, 
obtaining the parameters
$P_0=16.6$, $\alpha=1.95$ and $k^*= 0.14$, corresponding to 
$j^*=64,$ and a 
sample mean variance estimate $\sigma_{\bar{x}}^2=P_0/N=0.0055$.
The discrete power spectra ${\hat P}_j$ are then evaluated and 
averaged over a large number of long chains to measure the 
exact spectrum $P(k)$. 
The lower panel of Fig. \ref{fit} compares the fit of a single 
chain to this exact power spectrum.

\begin{figure}
\hskip -0.15in
\epsfig{file=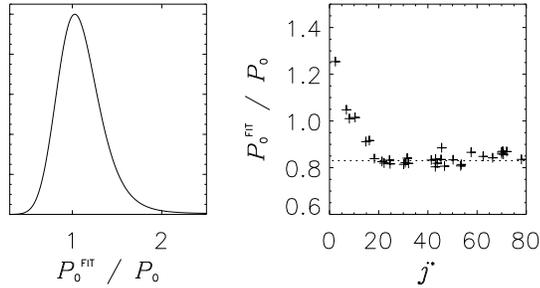,width=8.5cm,height=4cm}
\caption{(Left) Distribution for the quality of fit
$P_0^{fit}/P_0$, obtaining $P_0^{fit}$ by 
fitting 5000 individual chains, compared to the true value $P_0$ obtained by 
averaging. 
(Right) Quality of fit $P_0^{fit}/P_0$ at the lower $1\sigma$ level 
as a function of $j^*$, the number of Fourier modes in the white noise regime.
The dashed line indicates the mean value for $j^*>20,$ with very short 
chains tending to overestimate $P_0.$
}
\label{param_fit}
\end{figure}

Since $P_0, \alpha$ and $k^*$ are inferred from 
a single finite-length chain, 
not necessarily very far into the white noise regime, 
errors will be introduced in the sample best-fit parameters 
compared to the ideal fit to an infinitely long chain. 
These errors can be estimated by simulating a large 
number of finite chains, 
and fitting the template to each individual spectrum to measure 
the dispersion of the best-fit parameters.
We wish to avoid underestimating $P_0$, in order to prevent
diagnosing convergence too early.
For 5000 chains of length $N=3000$, 
we find $P_0^{fit}= 17 \pm 4$ (quoting the median and 68\% confidence 
limits). The exact value obtained from the averages of
all the parallel chains is $P_0=16.$ In
Fig. \ref {param_fit} we show the 
distribution of $P^{fit}_0/P_0$, with a lower $1\sigma$ (16th percentile) 
limit of 0.8. 
The accuracy of the template fit for varying chain lengths is then 
checked by 
measuring how $P^{fit}_0/P_0$ 
(at the lower $1\sigma$ level) varies as a function of the 
number of Fourier modes in the white noise regime $j^*,$ plotted in 
the left panel of Fig. \ref{param_fit}. 
For $j^* \gtorder 20,$ the estimate of $P_0$ has little scatter. 
For smaller $j^*$ the tendency is to overestimate $P_0.$

\begin{figure}
\hskip +0.35in
\epsfig{file=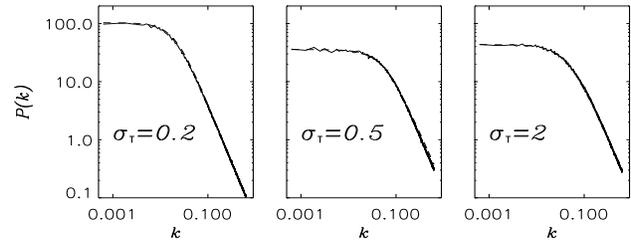,width=7.5cm,height=3cm}
\vskip +0.2in
\caption{True (solid) and fitted (dashed) power spectra of chains 
sampling the distribution as in Fig.~\ref{model}, using trial 
distributions of widths 
$\sigma_T/\sigma_0 =0.2$ (left), $0.5$ (middle) and  $2$ (right).}
\label{vary_size}
\end{figure}

We now investigate the quality of our template fitting procedure 
when the width $\sigma _T$ of the trial distribution is sub-optimal.
We test the goodness of fit of the template to the true power 
spectrum, as well as obtaining estimates for the possible errors on the 
best-fit parameters. 
The procedure described above is repeated with the same underlying 
distribution, but sampling with a trial distribution of varying width.
In Fig. \ref{vary_size} we show the averaged 
power spectra for chains generated using the range of step-sizes
$\sigma_T/\sigma_0 = 0.2, 0.5, 2$, with a best-fit template for 
comparison. 
Their power spectra all have the same form and the template fits well,
with the dispersion of the best-fit parameters given in Table \ref{table1}.
The possible underestimate for $P_0$ is in the range 
$ 0.7 < P_0^{fit}/P_0 <0.8$ at the 
$1\sigma$ level, which is suitably accurate for our purposes.
The slope of the curve at small-scales is 
given consistently by $\alpha \sim 2$, corresponding to near 
random-walk behaviour. 

\begin{table}
\begin{center}
\begin{tabular}{ccccc}
\hline\hline
$\sigma _T/\sigma _0$ & $P_0^{fit}$  & $\alpha$  &  $k^*$ &$P_0^{true}$\\\hline
$0.2$       & $110 \pm 30$ & $1.98 \pm 0.07$  & $0.019 \pm 0.005$ & $110$  \\
$0.5$       & $35 \pm 7$ & $1.97 \pm 0.09$ &  $0.057\pm 0.009$ & $35$ \\
$1.1$       & $17 \pm 4$ & $1.95 \pm 0.1$ &  $0.12 \pm 0.03 $ & $16$ \\
$2.0$       & $41 \pm 9$ & $1.90 \pm 0.10$   &  $0.05\pm 0.01$ & $43$ \\
\hline
\end{tabular}
\end{center}
\caption{Quality of template fit for chains sampling with various 
step sizes $\sigma _T/\sigma_0$. The distributions are obtained by 
fitting the power spectra of multiple chains to derive the 
median and 68\% limits for the best-fit variables $\ln P_0, \alpha$ and 
$\ln k^*$. We give the physical quantities 
$P_0$, $\alpha$ and $k^*$, with the true $P_0$ obtained by averaging for comparison.}
\label{table1}
\end{table}

\subsection{Testing for Convergence}

Any finite chain can be tested for convergence once the variables 
$P_0, \alpha$ and $k^*$ are obtained for each parameter separately. 
The following two requirements are made:

\vskip 4pt
\noindent
1. $k_{min}$ must be in the white noise regime $P(k) \sim k^0$,
defined concretely by the requirement $j^* > 20$. This insures that the 
correlated points are not biasing the distribution and indicates that the 
chain is drawing points throughout the full region of high probability.

\vskip 4pt
\noindent
2. The convergence ratio $r = \sigma^2_{{\bar x}}/ \sigma^2_0$ is 
calculated using the estimate for $P_0$, with $r=P_0/N$ for 
chains normalized to have unit variance. To obtain statistics with good 
accuracy, 
$r < 0.01$.

\vskip 4pt
\noindent
As an example Fig. \ref{fit} shows the spectra of a chain sampling the 
model distribution 
discussed in the previous section at three stages. 
After 500 steps the chain is still correlated at the 
largest scales measured and 
has not yet visited the entire distribution. 
It fails the convergence 
test immediately since $P(k)$ is not constant at large scales, with $j^*=5$.
After 1200 steps the chain has entered the 
white-noise regime at large scales with best-fitting $j^*=20$, but has 
not yet drawn enough samples to get good statistics; $r =0.02$ and 
the test is failed.
Finally after $N=2500$ steps the power spectrum is white noise at large
 scales, 
with $j^*=54$. The best-fitting $P_0=17.5$ gives a convergence ratio 
$r=0.007$, and the test is passed.

\begin{figure}
\hskip 0.4in
\epsfig{file=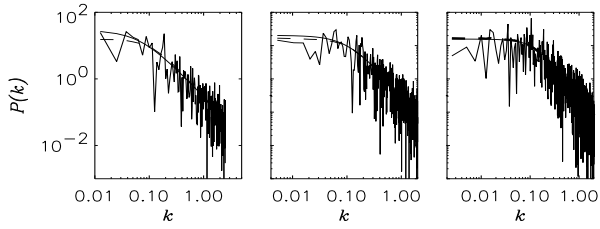,width=7.cm,height=3cm}
\caption{Discrete power spectra, with the best-fit template (solid) and 
true (dashed) power spectra for a chain of varying length
$N=500, 1200, 2500, $ from unconverged (left) to fully converged (right).}
\label{fit:b}
\end{figure}

\begin{figure}
\hskip 0.35in
\epsfig{file=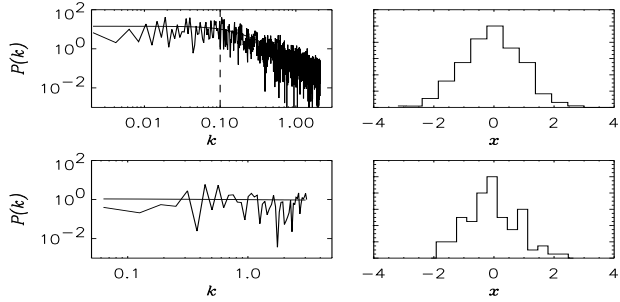,width=7.5cm,height=4cm}
\caption{Power spectrum (left) and resulting histogram (right)
of a chain sampling a model Gaussian distribution (top),
compared to the spectrum and histogram of the thinned chain (bottom)
where $k_t=0.1.$}
\label{thin}
\end{figure}

Once all the parameters have passed the test we are confident that 
the chain has converged and 
could stop it, but may choose to run it for longer to get more samples to 
reduce shot noise from the histograms. This is particularly 
relevant for lower 
dimensions where the chain can converge after relatively few steps. In 
practice we find that for high dimensional chains ($D\gtorder 8$), 
there are enough samples by the time the test is passed.
This would not be the case if we thinned the chain (i.e., saving only 
every $m$th point in the chain). This corresponds to cutting out 
small scale correlations with $k > k_t$, where $k_t= \pi/m$. The effect 
of thinning on the power spectrum and the 
recovered distribution is shown in Fig.~\ref{thin}. Not only is 
the sample mean variance increased by thinning, but the loss of points is very
noticeable in the shot noise of the histograms, particularly in higher 
dimensions. Since the convergence test assures us that the correlated points 
are not biasing the chain output, they are not removed.

\section{Optimizing efficiency and predicting convergence lengths}
\label{sec:optimal}

Because of limited computational resources, it is of great importance
to maximize the efficiency of an MCMC chain through a well chosen
trial distribution for the attempted steps. As discussed previously,
too small trial steps lead to random walk behaviour, leading to large 
inter-step correlations and hence slow convergence, whereas 
with too large trial steps jumps occur too infrequently, likewise
leading to large correlations and slow convergence. In this section 
we investigate how to choose an optimal trial distribution
between these two extremes.

The efficiency of a chain may be quantified as follows. Let
$\sigma^2 _{\bar x}(N)$ be the variance of the sample mean from
chains of length $N.$
The dimensionless efficiency $E$ of an MCMC chain is defined 
(see e.g. Neal 1992) by comparing its sample 
mean variance to an ideal chain (i.e., completely
uncorrelated) in the long chain limit:
\be
E=\lim _{N\to \infty } \frac{\sigma _0^2/N}{\sigma_{\bar x}^2(N)}.
\ee
Here $\sigma _0^2$ is the variance of the underlying distribution, and
$\sigma _0^2/N$ gives the sample variance of an ideal chain of length $N.$
$E^{-1}$ therefore gives the factor by which the MCMC chain is longer than 
an ideal chain yielding the same performance. A chain closest to ideal will 
have minimum correlation and therefore maximum efficiency.
The efficiency is related trivially to the number of steps $N_c$
needed to give a convergence ratio of $r$ at $N_c=(rE)^{-1}$ steps, since 
$r=\sigma_{\bar x}^2(N)/\sigma_0^2$. It follows that $E=\sigma _0^2/P(k=0).$
Previous work 
on choosing an optimal step-size for the trial distribution
includes Gelman, Roberts \& Gilks (1996) and Hansen \& 
Cunningham (1998), who investigate optimizing the efficiency for 
multivariate Gaussian underlying distributions. 

\begin{figure}
\hskip 0.3in
\epsfig{file=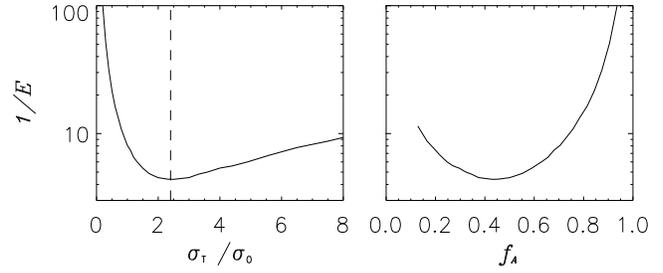,width=7.7cm,height=4cm}
\caption{Dependence of the inverse efficiency $1/E$ on the trial distribution
step-size $\sigma_T/\sigma_0$ (left) and the acceptance 
fraction $f_A$ (right) for a chain sampling 
a one-dimensional Gaussian model.}
\label{eff1d}
\end{figure}

We first consider how best to sample an underlying
$D$-dimensional Gaussian distribution
\ba
p({\bmath x})={1\over (2\pi {\sigma _0}^2)^{D/2}}\exp 
[-{\bmath x}^2/2\sigma _0^2]
\label{yyy}
\ea
where ${\bmath x}=(x_1,\ldots ,x_D), $ choosing a trial distribution
\ba
q({\bmath y})={1\over (2\pi {\sigma _T}^2)^{D/2}}\exp 
[-{\bmath y}^2/2\sigma _T^2].
\ea
We calculate the efficiency as a function of $\sigma_T$ by
running $10^4$ independent chains of length $N$
started with a point chosen at random according to $p({\bmath x}).$
For $D=1$ optimal efficiency is attained at 
$\sigma_T/\sigma_0=2.4$ (see Fig. \ref{eff1d}) 
at a maximum efficiency of $0.22$, giving a 
convergence length $N_c \approx 450$ for $r=0.01.$
The fraction of steps accepted is $f_A \approx 0.4$. 
In the diffusive, random walk regime, where $(\sigma_T/\sigma_0)\ll 1,$
almost all trial steps are accepted and the inverse efficiency scales as
$1/E \propto (\sigma_T/\sigma_0)^{-2}.$ 
At the other extreme $(\sigma_T/\sigma_0)\gg 1,$
almost no steps are accepted and the 
inverse efficiency scales as 
$1/E \propto \sigma_T/\sigma_0$,
which is also proportional to the acceptance rate.  
For $D=1,$ it is clearly better to err 
on the side of too large trial steps.

\begin{figure}
\epsfig{file=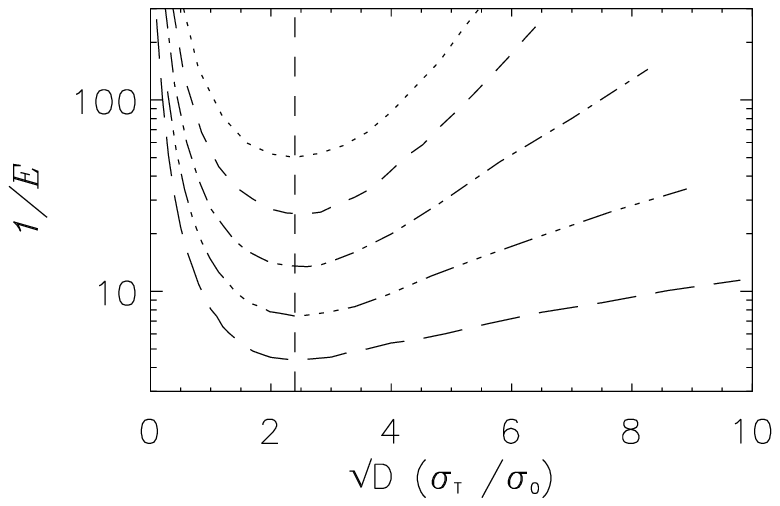,width=8cm,height=5cm}
\epsfig{file=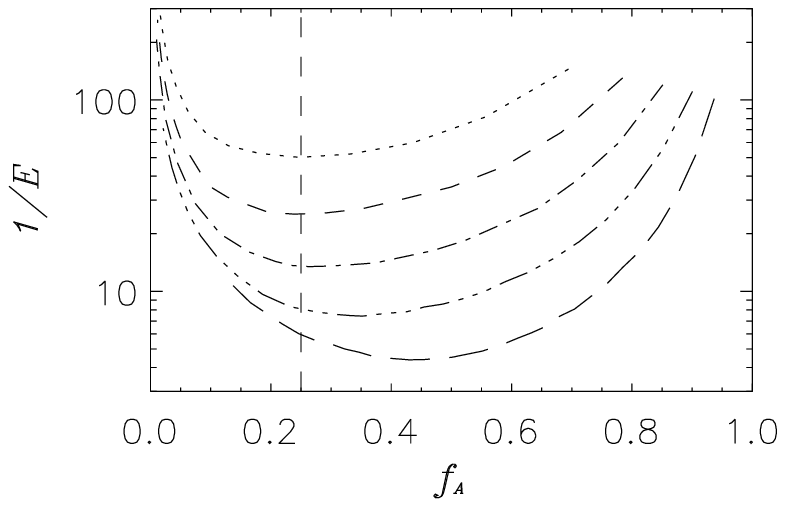,width=8cm,height=5cm}
\caption{Dependence of the inverse efficiency $E$ on the trial distribution
step-size $\sqrt{D}\sigma_T/\sigma_0$ (left) and the acceptance 
fraction $f_A$ (right) for a chain sampling 
a $D=1$ (bottom), $2, 4, 8$ and $16$ (top) dimensional Gaussian model.}
\label{ndeff}
\end{figure}

\begin{figure}
\epsfig{file=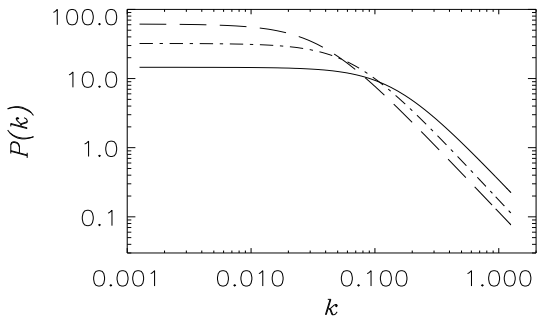,width=8cm,height=5cm}
\caption{Variation in power spectra with dimension 
for 4 (bottom), 8 and 12 (top) dimensional toy Gaussian models, sampled 
with a chain of length $N=5000$ using the optimal step-sizes obtained 
from Fig.~\ref{ndeff}.}
\label{diff_dim}
\end{figure}

For higher dimensions $(D>1)$ the results are shown in 
Fig.~ \ref{ndeff}.
The optimal efficiency follows a power law 
\be
E \approx {1\over 3.3 D},  
\ee
which works reasonably well for all $D> 1$ at an optimal step-size 
\be
\sigma_T/\sigma_0\approx 2.4/\sqrt{D}.
\ee
This translates into an optimal convergence length (for $r=0.01$)
of $N_c \approx 330~D$. 
We also found, as in Gelman et al.~(1996), that the acceptance 
rate of the optimal chain tends to 
$f_A=0.25$ for $D$ greater than about four.
We also observe that in higher dimensions it becomes progressively
more important to choose $\sigma _T$ not to large, because the efficiency 
for large steps scales as $E\propto (\sigma_0/\sigma _T)^D,$
as illustrated in Fig.~ \ref{ndeff}.
The theoretical notion described above that 
MCMC chain length scales linearly 
with dimension is only true for optimal $\sigma _T,$ 
which is much harder to achieve in high dimensions.
Fig. \ref{diff_dim} 
shows the power spectra for optimal chains in $4, 8$ and $12$ 
dimensions. It is clear that in higher dimensions the correlation 
length increases, seen as a lower wavenumber $k^*$ at which 
scale-free behaviour begins and leading to a higher $P_0$ and reduced 
efficiency.

In the general multivariate case, where 
\ba
p({\bmath x})={1\over (2\pi )^{D/2}}~{\rm det}^{-1/2}[{\bmath C}]~
\exp \left[-{1\over 2}{\bmath x}^T\cdot {\bmath C}^{-1}\cdot {\bmath x} \right]
\ea
where ${\bmath C}$ is the covariance matrix, choosing an acceptable
trial distribution amounts to choosing the $D(D+1)/2$ independent
elements of the covariance matrix for the trial distribution
${\bmath C}_T$ with sufficient accuracy. From simple rescaling, 
it is apparent that ${\bmath C}_T=(\sigma _T/\sigma _0)^2{\bmath C}$
is the optimal choice of trial distribution 
where $(\sigma _T/\sigma _0)$ is chosen as for the previous 
special case where ${\bmath C}=\sigma _0^2{\bmath I}.$

Fig. \ref{ndeff} shows how the efficiency varies when 
${\bmath C}_T \propto {\bmath C}$ but a sub-optimal scaling factor 
$\sigma _T/\sigma _0$ 
is used. It is quite likely that the two distributions will not be 
proportional, and in this case we can expect a great reduction in the 
efficiency. 
To illustrate the scenario in two dimensions we take a univariate 
trial distribution ${\bmath C}_T$ to sample a 
bivariate underlying distribution with covariance ${\bmath C}.$
If this distribution has widths differing by a factor of ten (e.g. by 
taking ${\bmath C}_{11}=1, {\bmath C}_{22}=100$), 
the optimum inverse efficiency attainable is $1/E=50$, 
found by varying $\sigma_T$ and calculating the efficiency as described 
earlier. It compares
badly with the optimal $1/E \sim 7.4$ when 
${\bmath C}_T \propto {\bmath C}$, and the chain takes seven times 
longer to converge than if a proportional trial distribution were 
used. In higher dimensions, sampling correlated underlying distributions with 
parameters a few orders of magnitude apart, the cost of using a 
non-optimal trial distribution can be even more extreme. 

Without prior knowledge of the underlying covariance $\bmath C,$ at least 
one preliminary chain will be necessary to obtain a reasonable 
estimate for the trial distribution. The number of steps required 
for such a chain 
can be estimated by first considering an uncorrelated chain, where points are 
drawn at random from a Gaussian distribution. 
To obtain its covariance matrix with sufficient accuracy we would like 
the square root of all its eigenvalues to be within 
$\sim 25\%$ of their true values. By varying the lengths of such 
chains in various
dimensions we find the length $N_{ideal}$ where this criterion is satisfied,
shown in table \ref{table2}.
For an MCMC chain with efficiency $E,$ a conservative estimate for the 
number of steps needed is then approximately $N_{mcmc}\approx N/E.$ 
These chain lengths 
are given in table \ref{table2} for optimal efficiency $E\approx 3.3D.$
Preliminary chains with very low efficiency would need to be 
significantly longer, so an iterative method to improve the covariance 
matrix estimate may be used. 

\begin{table}
\begin{center}
\begin{tabular}{ccc}
\hline\hline
  $D$    & $N_{ideal}$   &  $N_{mcmc}$   \\ \hline
 $4$     & $40$        &  $530$       \\
 $8$      & $110$       &  $2900$      \\
$16$     & $230$       &  $12000$     \\ 
\hline
\end{tabular}
\end{center}
\caption{Number of steps needed to estimate the
covariance matrix of a D-dimensional Gaussian distribution, 
such that the square root of all eigenvalues are correct 
to within 25\%. $N_{ideal}$ corresponds to an uncorrelated sampler; 
$N_{mcmc}=N_{ideal}/E$ provides an estimate for an MCMC chain 
with inverse efficiency 
$1/E \sim 3.3D$.}
\label{table2}
\end{table}


So far we have considered the chain length $N_c$ needed to satisfy the 
convergence criterion $\sigma^2_{\bar x} < 0.01 \sigma^2_0$.
We are often more interested however in limiting the extremes of a 
distribution, determining what part of the parameter space 
has been ruled out say at 2 or  $3\sigma$. Here `shot noise' 
or Poisson counting statistics becomes the primary limiting factor in
determining the fraction of points (and hence integrated probability) 
with probability $p < p_c$, when $p_c$ is small. 
If there were on average $n$ points in the region $p(x) < p_c$ 
(for $n \ll N$), the 
standard deviation in the count fluctuation would be $\sigma_n=\sqrt{n}$, 
giving a fractional error $\delta n /n = 1/\sqrt{n}$. 
This is true for an uncorrelated chain.
For an MCMC chain of length $N$ with efficiency $E$, the equivalent 
number of independent points is approximately $EN$, so a conservative 
estimate of the 
fractional error becomes $\delta n /n = 1/\sqrt{En}.$ 
To achieve a given accuracy requires a chain with $1/E$ more points 
than an uncorrelated chain.
We test that this expression gives an upper bound on the fractional 
error by studying the fluctuation in the number of counts obtained  
beyond both the 2 and $3 \sigma$ limits for a Gaussian model of 
dimension $D$, using multiple chains. We find 
$\delta n /n < 1/\sqrt{En}$ to hold for a large range of chain lengths 
over the range $1< D < 16$.

\section{What can go wrong: the worst case scenario}

To apply these methods to practical parameter estimation, we
must know how the power spectrum shape and the relation linking 
the covariance of the underlying distribution to that of the 
trial distribution for the optimal chain 
are altered for non-Gaussian 
underlying probability distributions
models, and whether it is possible for the convergence test to fail.
Clearly examples can be constructed where the power spectral
convergence test described above (as well as competing tests)
give the appearance that the underlying distribution has been
adequately sampled when this is not at all the case.
We consider a range of possible distributions, from mildly
non-Gaussian to more pathological distributions.

\subsection{Mild non-Gaussianity}

We consider two non-Gaussian models with 
probability densities 
\be
p_1(x)=\frac{1}{2}\bigl[p(x, \sigma_0=1) + p(x, \sigma_0=2)\bigr]
\ee
\be
p_2(x)=\frac{1}{2}\bigl[p(x, \sigma_0=1) + p(x, \sigma_0=4) \bigr]
\ee 
where $p(x, \sigma_0 )$ is the $D$-dimensional Gaussian in 
eqn.~(\ref{yyy}).
We investigate whether 
the power spectra of the resulting chains 
are well fit by our template
and how the optimal trial step-size relation 
is altered.
Both two- and eight-dimensional models are considered to include 
possible high-dimensional effects, and the 
efficiency is calculated as a function of step-size, 
$\sqrt{D}\sigma_T/\sigma_0.$

\begin{figure}
\hskip 0.15in
\epsfig{file=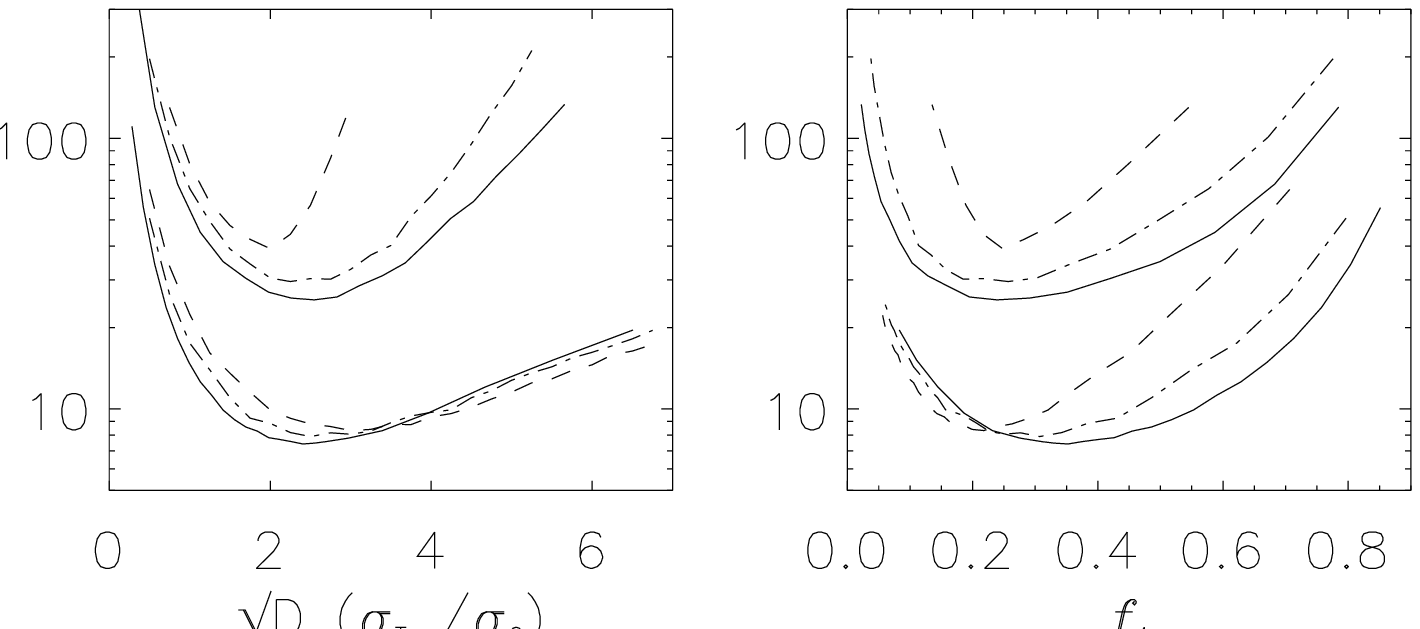,width=8.cm,height=4cm}
\begin{center}
\epsfig{file=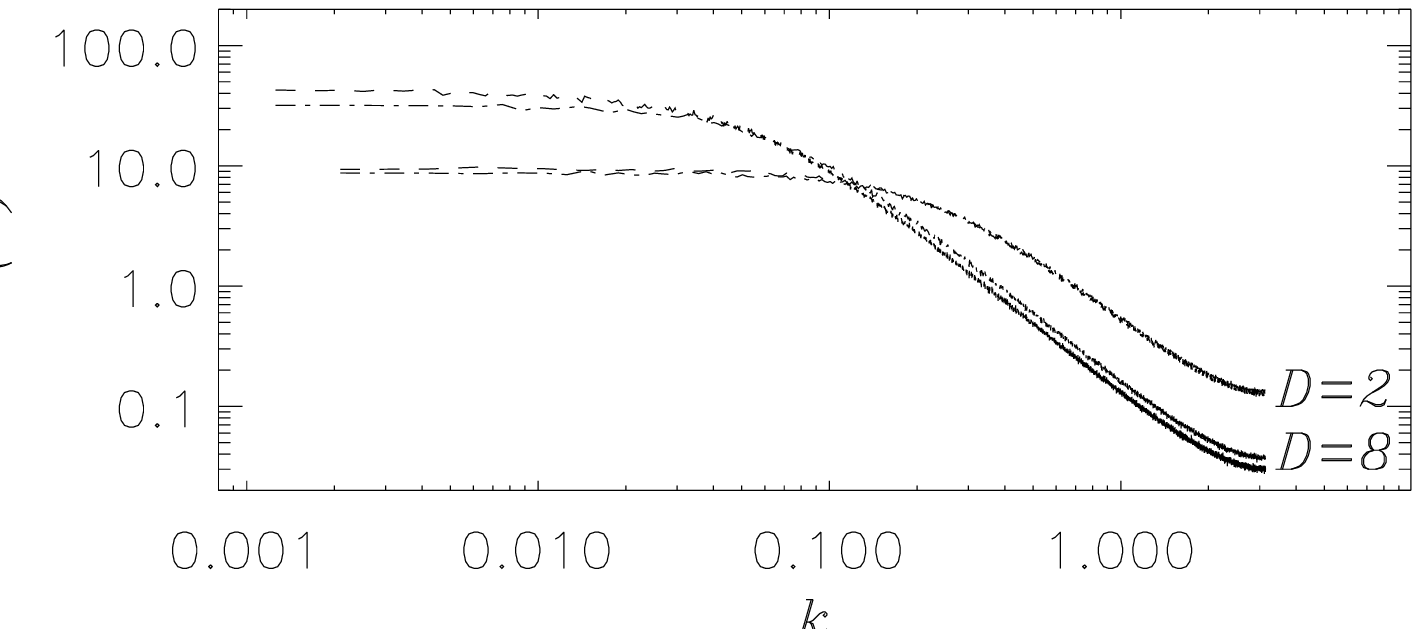,width=7.cm,height=3.5cm}
\end{center}
\caption{(Top) Dependence of the inverse efficiency $1/E$ on the 
trial step-size $\sigma_T/\sigma_0$ 
and acceptance rate $f_A$ for two non-Gaussian distributions, 
$p_1(x)$ (dot-dashed) and $p_2(x)$ (dashed), compared 
to the Gaussian model (solid) in two dimensions 
(lower curves) and eight dimensions (upper curves).
(Bottom) Smoothed power spectra for  
chains sampling the distributions $p_1(x)$ (dot-dashed) and $p_2(x)$ (dashed), 
in $D=2$ and $D=8$ dimensions.}
\label{ng_mild}
\end{figure}

Fig. \ref{ng_mild} shows the dependence of the inverse 
efficiency on the trial step-size and the 
acceptance fraction $f_A$, for both distributions $p_1(x)$ and $p_2(x).$ 
In two dimensions, little difference is observed in the behaviour for either 
distribution compared to the Gaussian case shown in solid lines. 
The optimal step-size for the underlying distribution $p_2(x)$ is slightly
higher, with a correspondingly lower acceptance rate. The 
efficiencies are very 
similar, although a chain with high acceptance fraction 
(with a smaller step-size) is more strongly penalised.
In eight dimensions the underlying distribution $p_1(x)$ 
shows similar behaviour to the Gaussian 
with optimal efficiency reduced by a factor 
$\sim 1.2$, but the more non-Gaussian $p_2(x)$
shows more extreme behaviour, with the inverse efficiency 
tightly peaked around the optimal step-size, with an 
efficiency lower by a factor of $\sim 2$ 
and a significant penalty for non-optimal sampling.
The choice of  $\sigma_T/\sigma_0=2.4/\sqrt{D}$ that one might make 
without prior knowledge of the form of these distributions would 
be too large for this eight dimensional case, but still
produces reasonably efficient chains for both 
distributions considered.
In the lower panels of Fig. \ref{ng_mild} the smoothed power spectra 
for both distributions are shown, displaying the same spectral behaviour
as the Gaussian chains.

\subsection{Curved distributions}

For $D\ge 2$ a potential problem arises when the region
of high probability is elongated and curved. Consider for
example the two-dimensional distribution
\ba
p(x_1,x_2)=A~\exp \left[
-{({x_1}^2+{x_2}^2-1)^2\over 8\sigma _r^2}
\right]
\ea
where the normalization $A$ is chosen appropriately and 
$\sigma _r\ll 1,$ or the more realistic non-symmetric
distribution 
\ba
p(x_1,x_2)=
A \exp \left[
-{({x_1}^2+{x_2}^2-1)^2\over 8\sigma _r^2}
-{x_2^2\over 2\sigma _\theta ^2},
\right]
\label{eqn:banana}
\ea
where $\sigma _\theta \ll 1$ but $\sigma _r < \sigma _\theta ^2.$ 
The first distribution is concentrated over an annulus of 
width $\sigma _r$ around the unit circle. To traverse the region of high
probability rapidly, one would want to attempt small steps
in the radial direction but comparatively larger 
steps in the azimuthal direction. Locally, this could
be achieved by a elongated bivariate Gaussian in the 
coordinates $x_1, x_2;$ however,
as one moves around, the axis of elongation would have to
rotate. However, in order to satisfy detailed balance, so that
the chain reproduces the underlying probability distribution,
it is necessary that $q_{trial}(x_{trial},x_i)$ remain 
constant throughout the simulation;\footnote{Clearly a
more complicated form for 
$q_{trial}({\bmath x}_{trial},{\bmath x}_i)$ not simply depending
on the difference $({\bmath x}_{trial}-{\bmath x}_i)$ can alleviate this 
slowdown, but for the general case
finding an appropriate distribution 
$q_{trial}$ may be difficult.}
therefore, an adaptive algorithm that allows $C_T$ to evolve
during the simulation is excluded. 

\begin{figure}
\hskip 0.1in
\begin{center}
\epsfig{file=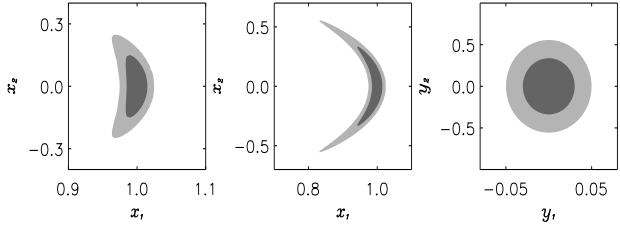,width=8.2cm,height=3.cm}
\end{center}
\caption{Two-dimensional probability distribution given 
by eqn.~(\ref{eqn:banana}) with variables $x_1, x_2,$ for 
$\eta=1$ (left), $\eta=5$ (centre),
and (right) the same distribution with transformed variables 
$y_1=x_1^2+x_2^2-1, y_2=x_2.$}
\label{crescent}
\end{figure} 

\begin{figure}
\hskip 0.25in
\epsfig{file=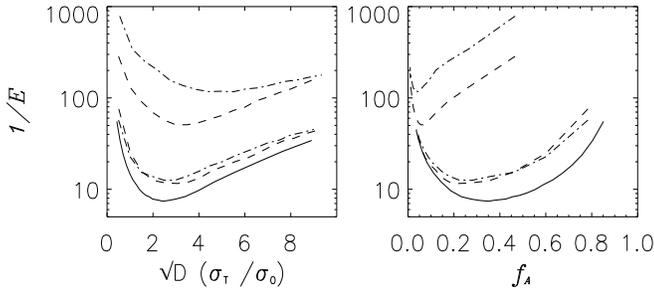,width=8cm,height=4cm}
\caption{Variation in inverse efficiency $1/E$ with step-size 
$\sigma_T/\sigma_0$, for the distributions shown in 
Fig. ~\ref{crescent} with curvature parameters $\eta=1$ and  
$\eta=5$ (upper curves) as defined in the text. Parameter $x_1$ 
is dot-dashed, $x_2$ is dashed, compared to the transformed parameters 
$y_i$ (solid).}
\label{curve_dist}
\end{figure} 

In the second example, the simply connected 
region of high probability has the shape of a thin crescent 
(shown in Fig. \ref{crescent}).
A simple change of variable can in principle solve the
problems described above. The pathology of this
distribution can be characterized by the dimensionless
parameter $\eta=\sigma ^2_\theta /\sigma _r,$ which gives the 
approximate number of standard deviations by which the 
region of highest probability is displaced from the 
two-dimensional mean. 
Any change of variable that renders the region convex 
(for example one which locally resembles polar coordinates)
makes the underlying distribution sufficiently close
to Gaussian in the new variable, so that the previously
described methods are adequate. Here for example the new parameters
$y_1=x_1^2+x_2^2-1, y_2=x_2$ transform the crescent to a bivariate
Gaussian. 

We investigate the effect of using the original coordinates 
on the efficiency of a chain sampled using the standard 
Gaussian trial distribution. We take two examples
with $\eta=1$ and $\eta=5$, whose distributions are
shown in Fig. \ref{crescent}, following the same procedure as described 
in section 5 to measure the efficiency as a function 
of the step-size. As
the distributions are no longer symmetric we use 
${\bmath C}_T=\sigma_T/\sigma_0 {\bmath C},$ measuring a separate efficiency 
for each 
parameter. In Fig. \ref{curve_dist} these results can be compared to
the optimal efficiency for a Gaussian, obtainable using the 
transformed coordinates (solid line).
For the mildly curved distribution with $\eta=1$ the efficiency is 
simply reduced by a factor $\sim 1.5$, but the optimal step-size is unchanged.
The more extreme $\eta=5$ case shows markedly different behaviour however,
with the less efficient $x_1$ parameter fifteen times slower than the
optimal Gaussian case, for a step-size twice as big and a far lower 
acceptance rate, $f_A \sim 0.05$. The high curvature and long tails make it 
impossible to converge quickly using a Gaussian trial distribution, since 
the trial and underlying distributions overlap so badly. Low acceptance 
rates combined with diffusive steps result in highly correlated chains. 
In practice this slowdown is readily diagnosed using our spectral test
while running the chain.
Coupled with a tendency towards a very low acceptance rate
for a naive optimal step-size $\sigma_T/\sigma_0 \sim 2.4/\sqrt{D},$ 
this would be a clear indication of the need to reparametrize.

In high dimensions it is often difficult to
identify an appropriate change of variables 
empirically, although Kosowsky, 
Milosavljevic \& Jimenez (2002) and Jimenez et al. (2004) 
describe such a transformation of a set of simple-inflationary
cosmological parameters into a set of nearly-orthogonal 
`physical' parameters.
In practice however, we have found the non-transformed
cosmological parameters to be sufficiently well-behaved
for flat cosmologies, achieving high efficiencies.

\subsection{Bimodal or multi-peaked distributions}

Failure of the spectral convergence test is most likely to 
occur for distributions having multiple narrow
peaks of high probability connected by wide regions of low
probability. While the length $N_{\rm single~peak}$
required to sample adequately the region around a single peak
may be rather short, the length $N_{\rm tunnel}$
required to tunnel to other peaks may be substantially
longer. 

The simplest example is a symmetric bimodal
distribution 
\ba
p(x)={1\over 2}{1\over \sqrt{2\pi }}\left\{
 \exp \left[ -{(x-a)^2\over 2}\right]
+\exp \left[ -{(x+a)^2\over 2}\right]
\right\}
\ea
with variable peak position at $x=\pm a$. We wish to know 
if the spectral convergence test can fail for this model by indicating 
convergence before the full distribution is recovered.
The behaviour of the chain will depend on both $a$ and the trial 
step-size $\sigma_T/\sigma_0$. We explore this dependency and its 
effect on the chain power spectrum by sampling from the 
distribution with variable peak separation $a$, using a fixed trial 
distribution width $\sigma_T$.

\begin{figure}
\hskip +0.1in
\epsfig{file=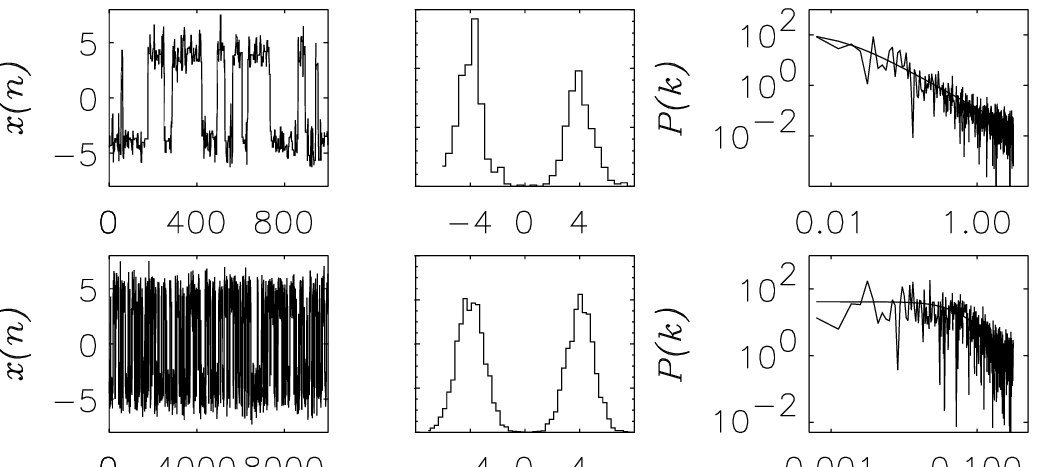,width=8.5cm,height=4.cm}
\vskip +0.3in
\hskip +0.1in
\epsfig{file=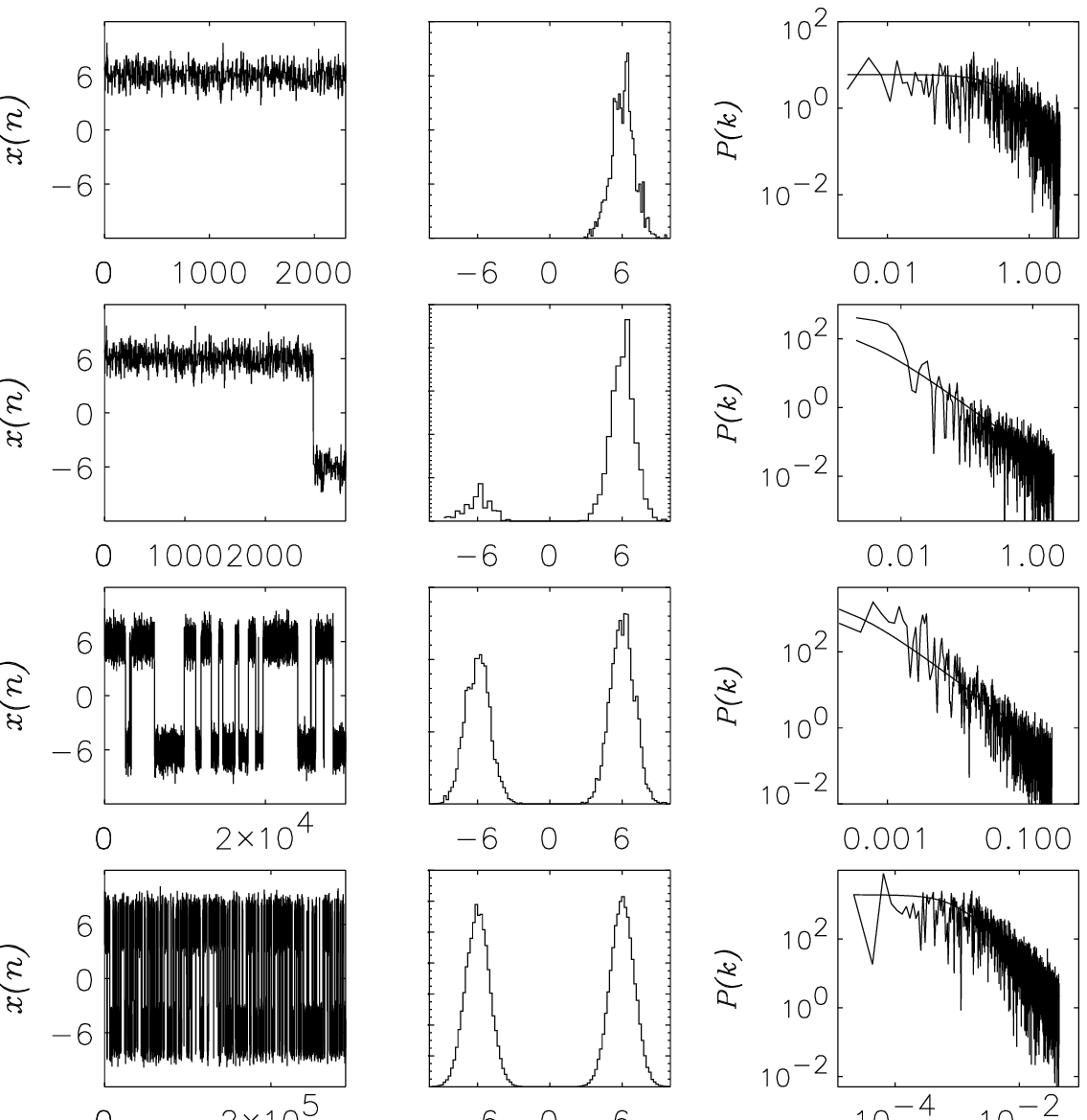,width=8.5cm,height=8.cm}
\vskip +0.3in
\hskip +0.1in
\epsfig{file=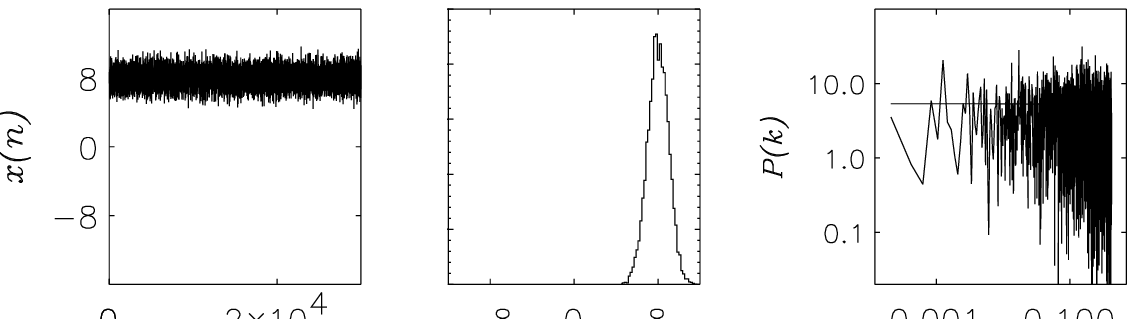,width=8.5cm,height=2cm}
\vskip +0.2in
\caption{Behaviour of chains with fixed trial distribution 
sampling a bimodal distribution with peak positions $x=\pm 4$ 
(top 2 panels) $x=\pm6$ (middle 4 panels) and $x=\pm 8$ (bottom panel). 
The chain output (left), recovered distributions (centre) and
power spectra (right) are shown for various chain lengths $n$.}
\label{bimod1}
\end{figure} 

The observed behaviour, shown in Fig.~\ref{bimod1}, can be 
split roughly into three categories.
For small peak separation $a$ the chain jumps between the two peaks 
frequently, converging toward both peaks simultaneously. The power spectrum 
only 
displays large-scale white-noise behaviour at late times when both peaks 
are fully sampled. The template fits the spectrum well at early and late times,
shown in Fig.~\ref{bimod1} (top two rows).
If $a$ is increased sufficiently, then the chain may 
converge `locally' on one of the peaks before it takes its 
first jump to the second peak, shown in the third row of 
Fig.~\ref{bimod1} for $a=6$. Before it makes this 
jumps the chain 
would pass the spectral convergence test, but as soon as it swaps 
peaks an excess of power is produced on large scales (4th row). It now 
fails the convergence test, although the template does not provide 
a very good fit. 
After a much longer time (6th row) the 
chain will have sampled both peaks sufficiently to converge `globally', 
recovering the correct distribution with peaks of equal height. 
The chain is no longer biased 
towards one of the peaks, so again the power becomes white-noise on 
large scales and the convergence test is passed. 
For large separation $a$ the chain can sample one peak without jumping 
to the second peak in any reasonable time. This is observed in 
Fig.~\ref{bimod1} (bottom row) for $a=8$. For a length ten times 
longer than $N_C$ it does not jump, and we would falsely conclude that 
the distribution had only one peak. 
The same effect is seen if the peak separation is narrower (e.g. $a=6$) 
but the trial step-size is very small.

From these models we can conclude that if 
the chain visits a second peak even briefly, 
then it will show up as excess large-scale power in the power spectrum 
and the chain will not pass the convergence test until it has sampled 
both peaks sufficiently. 
This is useful since in high dimensions 
a jump might not be so obvious in the time-ordered chain output. 
If the peaks are very widely separated (or if an overly 
small step-size is being used) then the chain might not leave the first 
peak in a reasonable time. If there are just two peaks then multiple 
chains may be used to diagnose this problem, 
but for multiple disconnected peaks this could be 
insufficient, although this is an unlikely scenario for the case 
of cosmological parameters.

\begin{figure}
\hskip 0.12in
\epsfig{file=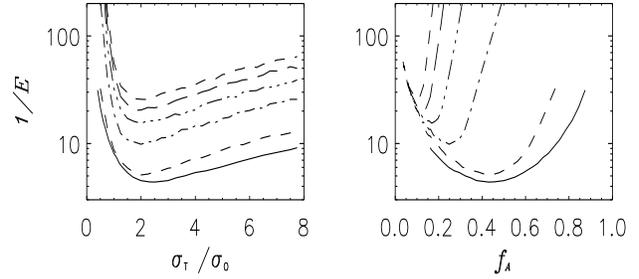,width=8.2cm,height=4.5cm}
\caption{Variation in inverse efficiency $1/E$ with trial 
step-size $\sigma_T/\sigma_0$ and acceptance rate $f_A$ 
for a one-dimensional bimodal distribution with Gaussian 
peaks at $x=\pm a$ for $a=0$ (bottom), $2, 4, 6, 8$ and $10$ (top).}
\label{bimod2}
\end{figure} 

Finally we investigate how the optimal scaling relations are altered if the 
distribution is bimodal, considering a range of peak separations from 
$a=2$ to $a=10$. Using the same methods described
earlier, we find that the optimal step-sizes are 
similar to the Gaussian case, shown in Fig.~\ref{bimod2}, although slightly 
reduced to $\sigma_T/\sigma_0 \sim 2.$  It is important to 
note that $\sigma_0$ applies here to the full distribution, not to a 
single peak.
The optimal acceptance rate decreases as the peak separation increases, 
to $f_A \sim 0.1$ for $a=10.$ The inverse 
efficiency then gradually increases as the chain becomes more correlated 
as a consequence of steps occurring more infrequently.
The unusual behaviour of the chain getting stuck indefinitely 
on one peak, seen in the final row of Fig. ~\ref{bimod1}, occurs 
for step-sizes smaller than those considered here.

\section{Sampling Method}

This section outlines an explicit sampling procedure based on the 
methods and considerations presented in the previous sections.
This is the procedure that we applied to extracting 
cosmological parameters from
the CMB+LSS data. 

\vskip 4pt
\noindent
1. An initial best guess is made for the covariance matrix of 
the underlying distribution ${\bmath C}_1,$ which is used
to fix the multivariate Gaussian trial
distribution characterized by the
covariance matrix ${\bmath C_T}_1$, chosen according to the rule
\ba
{\bmath C_T}_i=(2.4^2/D){\bmath C}_i.
\label{eqn:cov}
\ea
If possible, a starting point in the high likelihood region 
is chosen.

\vskip 4pt
\noindent
2. A short chain is sampled using ${\bmath C_T}_1$ to 
obtain a refined estimate for  ${\bmath C}_2,$ 
which in turn is used to update  ${\bmath C_T}$
according to the same rule in eqn.~\ref{eqn:cov}
If this first chain started in a region of low likelihood, the first 
section (where $L/L_{max} \ltorder 0.1$) is discarded before estimating 
${\bmath C}_2.$
If the acceptance rate for this initial chain is very low 
(e.g. $\ltorder 0.01$) or 
very high (e.g. $\gtorder 0.9$), then this first chain does not 
provide a useful estimate of the covariance matrix and a new guess 
should be made for ${\bmath C}_1$.

\vskip 4pt
\noindent
3. A second chain is started with trial distribution ${\bmath C_T}_2,$ 
starting where
the previous chain finished.
The process is iterated until further refinement would not lead to 
a sufficiently significant improvement in efficiency. 
The efficiency compared to `optimal' can be judged using both the 
acceptance rate as an indicator, and the power spectrum test 
to give an estimate for $P_0.$ 
For real, approximately Gaussian distributions tested for dimension 
$D\ltorder 8$ 
we have found that one update of the trial distribution 
to ${\bmath C_T}_2$ normally suffices to sample efficiently. 

\vskip 4pt
\noindent
4. Only the final chain is used for the analysis, which is
tested for convergence once $N>N_c$. 
The best-fit template for the power spectrum of each cosmological parameter
is found, to give
$P_0$, $\alpha$ and $k^*$. The test is passed if $k_{min}$ is in 
the regime $P(k) \propto k^0,$ defined by the condition $j^* >20,$ and 
if the convergence ratio $r < 0.01$ for each parameter.
If the white noise-regime has been reached but $r > 0.01$, an
estimate can be made for how much longer the chain needs to run  
using $r \propto 1/N$.

\vskip 4pt
\noindent
6. Once the chain has converged, the histograms 
of the number densities are checked to 
insure sufficient points have been collected, and further sampling 
may be carried out if a characterisation of the wings of the distribution
at high $\sigma$ are desired.

\vskip 4pt
\noindent
7. If a series of updates of the trial distribution fails to improve the 
efficiency or produce convergence, often signalled by a very low 
acceptance rate, it is likely that the distribution is highly non-Gaussian 
and reparametrization should be explored.

\section{Application to CMB Parameter Estimation}

\begin{figure}
\epsfig{file=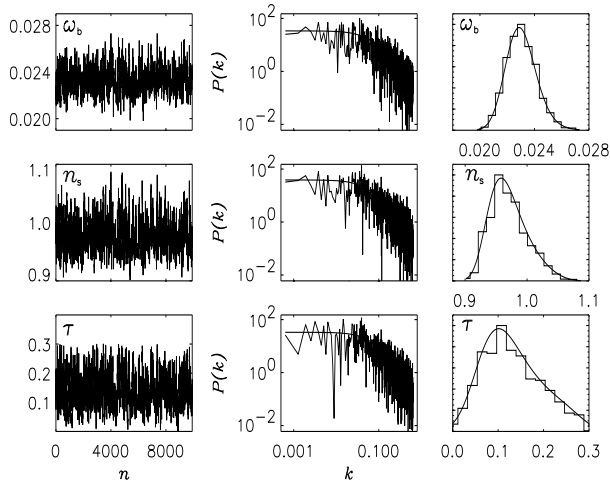,width=8.5cm,height=6cm}
\vskip 0.2in
\caption{MCMC chain for a pure adiabatic cosmological model, showing
the chain output (left), power spectra (middle) and marginalized distributions 
(right) for three parameters of 
a 7-dimensional model, sampled with a chain of length $N=9000$.}
\label{adiabatic}
\end{figure}

We applied the above methods to estimating cosmological parameters using the 
CMB and LSS data and code as described in Bucher et al. (2004). 
Initially we consider a 7-dimensional 
pure adiabatic cosmological model defined by the 
parameters $\omega_b, \omega_d, 
\Omega_\Lambda, n_s, \tau, \beta,$  and an overall amplitude $A_s$. 
An appropriate 
trial distribution and high-likelihood starting point are easy to 
obtain given already-available results for this distribution in 
Spergel et al. (2003). 
After updating the trial
distribution once, our second and final chain runs for 4000 steps with an 
acceptance rate of 
26\% before achieving convergence with white noise power at large 
scales and $r<0.01$ for all parameters. The values of the 
power spectrum template variables are
given in Table \ref{table3} for the most and least 
efficient parameters.
The underlying distribution is fairly close to Gaussian. 
We are able to sample 
very efficiently, achieving an inverse efficiency of 
$\approx 40$, compared to an optimal inverse efficiency of $\approx 25$ for 
$7$ dimensions. 
We continued running the chain to 9000 steps for improved statistics.
Fig. \ref{adiabatic} shows the time 
series of three of the seven parameters, their power spectra 
and best-fit curves and the recovered marginalized posterior 
distributions. 
The smoothed posterior distributions are obtained by fitting 
the logarithm of the binned number densities to a high order 
polynomial, as described in Tegmark et al. (2003).

\begin{figure}
\hskip 0.2in
\epsfig{file=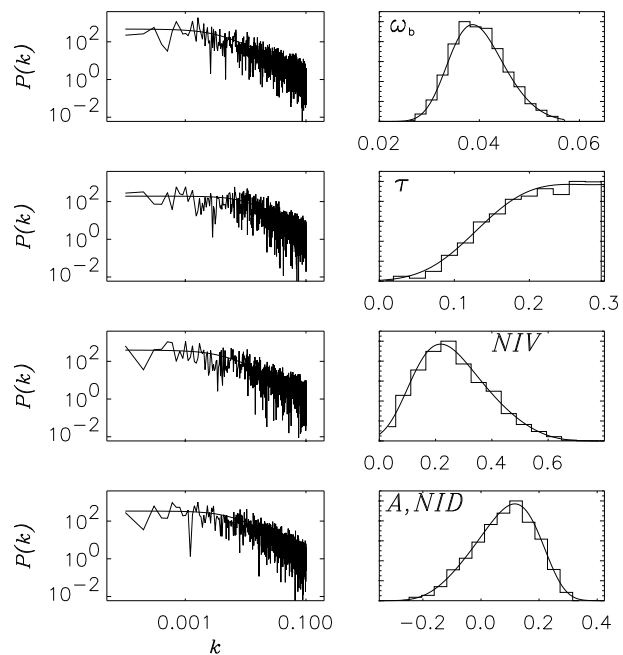,width=8.5cm,height=8.5cm}
\caption{
MCMC chain for a mixed adiabatic/isocurvature cosmological model.
Power spectra (left) and marginalized distributions (right) 
for four parameters of the 16-dimensional model: the baryon density 
$\omega_b$, optical depth $\tau$ and relative power in two 
non-adiabatic modes, neutrino velocity ($NIV$) and correlated 
adiabatic/neutrino density isocurvature ($A,NID$).}
\label{isocurv}
\end{figure}

We then extend the parameter space to 16 dimensions, including general
isocurvature perturbations, requiring nine extra parameters to quantify the 
additional mode contributions. As described in Bucher et al.~(2004), we 
include the 
CDM isocurvature mode and the neutrino density and velocity isocurvature 
modes, creating a $4\times4$ symmetric mode matrix with 
the adiabatic mode and cross-correlations. Physically all the 
eigenvalues of this matrix must be non-negative. This constraint was  
imposed by assigning a zero likelihood to those trial steps
violating it.
With very little prior knowledge of this probability distribution and a 
higher cost of sampling non-optimally, the trial distribution was 
updated more times to improve the efficiency. 
The chain converged after $6\times 10^4$
steps, with $r_i < 0.01$ for 
all parameters. In this chain 60\% 
of steps violated the constraint, so that only $N_{models}=2.4\times10^4$ 
actual cosmological models were evaluated.
In Fig. \ref{isocurv} we show power spectra and the recovered distributions 
for four of the parameters, with the template variables given in 
Table~\ref{table3}. 
By taking an effective number of steps 
$N_{models}=24000$ we find an inverse 
efficiency of $1/E=240$ for the worst parameter. 
Since this distribution is non-Gaussian and of high dimension, it is not 
surprising that we do not achieve the optimal Gaussian inverse 
efficiency of $1/E=53.$

\begin{table}
\begin{center}
\begin{tabular}{ccccc}
\hline\hline
    &AD  & AD   &  AD+ISO  & AD+ISO  \\ 
& best  &  worst  &   best & worst \\ \hline
$P_0=1/E$      & $33$   & $39$  & $185$  & $498$  \\
$\alpha$     & $1.8$   & $1.8$  & $1.8$  & $1.8$ \\
$j^*$        & $90$  & $77$  & $92$  & $37$ \\
\hline
\end{tabular}
\end{center}
\caption{
Best-fit template variables for the 
power spectra of chains sampling cosmological parameters. 
All chains are normalized to have unit variance. 
The template parameters for the best (i.e., lowest $1/E$) 
and worst cosmological parameters are indicated for both
the adiabatic models (AD) and the 
mixed adiabatic/isocurvature models (AD+ISO).}
\label{table3}
\end{table}

\section{Discussion}

We have shown how a spectral test based on an empirical
fitting function can be used to diagnose reliably the 
convergence properties of a single long MCMC chain. 
Explicit criteria for convergence using this test
were formulated and subsequently demonstrated to detect 
lack of convergence for a variety of underlying distributions.
A procedure to optimize the covariance matrix for a Gaussian
trial distribution was also explored. 

The distributions for which our test did not
successfully detect a lack of convergence 
were those with widely separated, narrow
multiple peaks of high probability density.
Although for such distributions no universally
valid procedure exists for uncovering the presence
of multiple peaks, a possible method is to run 
multiple chains starting from widely separated,
randomly chosen points in the sample space. 
The Gelman \& Rubin test
provides an appropriate convergence 
diagnostic for multiple chains. The variance
of the sample means of the several chains is compared to the 
estimated variance of the sampled distribution, 
obtained by averaging over the 
sample variance within each chain. 
In the case of multiple peaks or lack of convergence 
about a single peak, the variance of the sample means 
will be too high, failing to fall below a specified 
fraction of the within-chain sample variance.

The spectral test applied to single chains has 
the advantage of exploiting all the 
information relevant to estimating the
sample mean variance. Rather than 
simply comparing a small number of parallel chains,
the spectral test effectively divides
the chain in $2j_{max}$ independent ways when 
all the $\hat P_j$ with $1\le j\le j_{max}$
are used to fit to the template. 

We find that the spectral test and 
trial distributions optimized according
to the procedure outlined above work 
well when applied to CMB parameter
estimation. As described in detail
above, a chain of length $N=4\times 10^3$
suffices to obtain $r\ltorder 0.01$
(equivalent to a $10\% $ standard
deviation for the sample means of the 
cosmological parameters) 
for an adiabatic cosmological model,
although longer chains were actually
employed to reduce shot noise error
in the outer edge of
the region of high probability.
For the full adiabatic/isocurvature
model, convergence was attained within
$N=6\times 10^4$ steps, which 
involved calculating only $2.4\times 10^4$
distinct cosmological models due to large
zero-likelihood regions. Neither
model showed evidence of any of the 
pathologies explored in section 6. 
The chain performance attained may be 
compared to that expected from 
a Gaussian of the same dimension
explored using a chain with an
optimal trial distribution.
Our efficiencies were lower
by factors of approximately
$1.6$ and $5$ for the two
cases, respectively. 

\section*{Acknowledgements}
JD was supported by a PPARC studentship. 
MB thanks Mr D.~Avery for support through the 
SW Hawking Fellowship in
Mathematical Sciences. PGF thanks the Royal Society. 
KM was supported by a PPARC Fellowship and a Natal University research grant. 
CS was supported by a Leverhulme Foundation grant. 
We thank Martin Kunz, David Parkinson and David Spergel for 
useful discussions.


\label{lastpage}


\begin{thebibliography}{99}



\bibitem{christensen} Christensen~N., Meyer~R., Knox~L., Luey~B., 2001, 
Class.~Quant.~Grav., 18, 2677
\bibitem{coda} Best~N.~G., Cowles~M.~K.,Vines~K., 1995, Technical Report, 
Medical Research Council Biostatistics Unit, Cambridge
\bibitem{bdfms} Bucher~M., Dunkley~J., Ferreira~P.~G., Moodley~K., Skordis~C., 
2004, PRL, in press (astro-ph/0401417)
\bibitem{cowles} Cowles~M.~K., Carlin~B.~P., 1996, 
J.~Amer.~Statist.~Assoc, 91, 434
\bibitem{Doran} Doran~M., Mueller~C., 2003 (astro-ph/0311311)
\bibitem{Gel96}
Gelman~A., Roberts~G.~O., Gilks~W.~R, 1996, in eds
Bernardo~J.~M., Berger~J.~O., Dawid~A., Smith~A., {\em{Bayesian Statistics 5}}.
599, OUP 
\bibitem{Gelrub} Gelman~A., Rubin~D.~B., 1992, Statist. Sci., 7, 457
\bibitem{geweke} Geweke~J., 1992, in eds
Bernardo~J.~M., Berger~J.~O., Dawid~A., Smith~A., {\em{Bayesian Statistics 4}} 
169, OUP 
\bibitem{gilks} Gilks ~W.~R., Richardson~S., Spiegelhalter~D.~J., 1995, 
{\it Markov Chain Monte Carlo in Practice}. Chapman \& Hall, London 
\bibitem{Han98} Hanson~K.M., Cunningham~G.~S., 1998, Proc. SPIE,  3338, 371
\bibitem{heid} Heidelberger~P., Welch~P.~D., 1981, Comm.~ACM, 24, 233; 1985, 
Operations~Research, 31, 1109
\bibitem{wmap_hin} Hinshaw~G. \ETAL, 2003, ApJS, 148, 135  
\bibitem{jimenez} Jimenez~R., Verde~L., Peiris~H., 
Kosowsky~A., 2004, PRD, submitted (astro-ph/0404237)
\bibitem{wmap_kog} Kogut~A. \ETAL, 2003, ApJS, 148, 161
\bibitem{kosow} Kosowsky~A., Milosavljevic~M., Jimenez~R., 2002, 
PRD, 66, 063007
\bibitem{Knox} Knox~L., Christensen~N., Skordis~C., 2001, ApJ, 563, L95
\bibitem{Lew02} Lewis~A., Bridle~S., 2002, PRD, 66, 3511
\bibitem{metropolis} Metropolis~N., Rosenbluth~A.~W., Rosenbluth~M.~N., 
Teller~A.~H., 1953, J.~Chem.~Phys., 21, 1087 
\bibitem{neal} Neal~R.~M., 1993, Technical Report CRG-TR-93-1 
\bibitem{2dF} Percival~W.~J. \ETAL, 2001, MNRAS, 327, 1297
\bibitem{raftery} Raftery~A.E., Lewis~S.~M., 1992, in eds
Bernardo~J.~M., Berger~J.~O., Dawid~A., Smith~A., {\em{Bayesian Statistics 4}} 
763, OUP 
\bibitem{slosar} Slosar~A., Hobson~M.~P., 2003, MNRAS, submitted 
(astro-ph/0307219)
\bibitem{spergel} Spergel~D. \ETAL, 2003, ApJ, 148, 213
\bibitem{tegm} Tegmark~M. \ETAL, 2003, PRD, in press (astro-ph/0310723)
\bibitem{sdss_tegmark} Tegmark~M.\ETAL, 2003, ApJ, in press (astro-ph/0310725)
\bibitem{verde} Verde~L. \ETAL, 2003, ApJS, 148, 195

\end{thebibliography}
\end{document}